\documentclass[prd,aps,preprint,nofootinbib,superscriptaddress]{revtex4}
\usepackage{epsfig}
\usepackage{amsmath}

\begin{document}

\preprint{BNL-NT-05/26} \preprint{RBRC-537}

\title{Single-Transverse Spin Asymmetries: \\ From DIS to Hadronic
Collisions}

\author{Werner Vogelsang}
\email{vogelsan@quark.phy.bnl.gov}
\affiliation{Physics Department, Brookhaven National Laboratory,
Upton, NY 11973} \affiliation{RIKEN BNL Research Center, Building
510A, Brookhaven National Laboratory, Upton, NY 11973}
\author{Feng Yuan}
\email{fyuan@quark.phy.bnl.gov}
\affiliation{RIKEN BNL Research Center, Building 510A, Brookhaven
National Laboratory, Upton, NY 11973}
%\date{\today}
\begin{abstract}
We study single-spin asymmetries in semi-inclusive deep inelastic
scattering with transversely polarized target. Based on the QCD
factorization approach, we consider Sivers and Collins
contributions to the asymmetries. We fit simple parameterizations
for the Sivers and Collins functions to the recent HERMES data,
and compare to results from COMPASS. Using the fitted
parameterizations for the Sivers functions, we predict the single
transverse spin asymmetries for various processes in $pp$
collisions at RHIC, including the Drell-Yan process and angular
correlations in di-jet and jet-plus-photon production. These
asymmetries are found to be sizable at forward rapidities.

%\vspace{10cm}
\end{abstract}
\maketitle

\newcommand{\be}{\begin{equation}}
\newcommand{\ee}{\end{equation}}
\newcommand{\ben}{\[}
\newcommand{\een}{\]}
\newcommand{\beqn}{\begin{eqnarray}}
\newcommand{\eeqn}{\end{eqnarray}}
\newcommand{\Tr}{{\rm Tr} }

\section{Introduction}

Single-transverse spin asymmetries (SSA) in hadronic processes
have a long history, starting from the 1970s and 1980s when
surprisingly large SSAs were observed in $p_{\uparrow}p\rightarrow
\pi X$ \cite{E704} and $pp\to \Lambda_{\uparrow}X$ \cite{Bunce} at
forward rapidities of the produced hadron. They have again
attracted much interest in recent years from both experimental and
theoretical sides \cite{review}. In particular, first measurements
by the STAR, PHENIX, and BRAHMS collaborations at RHIC have now
become available \cite{STAR,PHENIX,BRAHMS}, which extend the SSA
observations from the fixed-target energy range to the collider
regime. Again, large asymmetries were found in
$p_{\uparrow}p\rightarrow \pi X$ at forward rapidities of the
produced pion. Meanwhile, experimental studies in Deep Inelastic
Scattering (DIS) by the HERMES collaboration at DESY, by SMC at
CERN, and by CLAS at the Jefferson Laboratory also show remarkably
large SSAs in semi-inclusive hadron production,
$\gamma^*p_{\uparrow} \rightarrow \pi X$
\cite{smc,hermes,HERMESt,HERMESnew,jlab}. Data from COMPASS for scattering
off deuterons have been published as well~\cite{COMPASS}, which show
no large asymmetry.  On the
theoretical side, there are several approaches to understanding
SSAs within Quantum Chromodynamics (QCD) \cite{review,Efremov,qiu}. Recent
interest focuses on the role of partonic transverse momentum in
creating the observed asymmetries. Transverse-momentum-dependent
(TMD) parton distributions and fragmentation functions, and their
relevance for semi-inclusive DIS (SIDIS), the Drell-Yan process,
and single-inclusive hadron production at hadron colliders have
been investigated in
\cite{RalSop79,ColSop81,ColSop81p,ColSopSte85,
Siv90,Col93,Kot94,Ans94,MulTan96,BroHwaSch02,Col02,BelJiYua02,BoeMulPij03}.
Compared to the normal integrated distributions, the TMD
distributions provide much more information; for example, some of
them contain information on orbital angular momenta of partons in
the nucleon and have also been linked to spatial distributions of
partons \cite{JiMaYu02,Burkardt}.

The Sivers function \cite{Siv90} is one of these interesting TMD
parton distributions. It represents a distribution of unpolarized
quarks in a transversely polarized nucleon, through a correlation
between the quark's transverse momentum $\vec{k}_\perp$ and the
nucleon polarization vector $\vec{S}_\perp$. The existence of the
Sivers function requires final/initial-state interactions, and an
interference between different helicity Fock states of the
nucleon. In the absence of interactions, the Sivers function would
vanish by time-reversal invariance of QCD, hence it is often
referred to as a ``naively time-reversal-odd'' distribution. As
was shown in \cite{BroHwaSch02,Col02,BelJiYua02}, the interactions
are represented in a natural way by the gauge link that is
required for a gauge-invariant definition of a TMD parton
distribution. Interference between different helicity Fock states
implies nonzero orbital angular momentum
\cite{BroHwaSch02,JiMaYu02}. Both these properties motivate the
study of this function. The Sivers function will contribute to the
target SSA in semi-inclusive DIS, but also to SSAs in polarized
$pp$ scattering processes such as the Drell-Yan process and di-jet
or jet-photon correlations. We will discuss all these asymmetries in this paper.

The Collins function is another ``naively time-reversal-odd''
function. It is a transverse-momentum dependent fragmentation
function and was introduced in \cite{Col93}. It represents a
correlation between the transverse spin of a fragmenting quark and
the transverse momentum of the hadron relative to the ``jet axis''
in the fragmentation process. Like the Sivers function, it
vanishes when integrated over all transverse momentum. Indications
of a nonvanishing Collins effect have been found in semi-inclusive
DIS \cite{HERMESt}. Very recently results for measurements in
$e^+e^-$ annihilation to two hadrons have been reported, which
give direct evidence for the Collins effect \cite{belle}.

The formulation and study of TMD functions is really useful only
when they appear in processes for which QCD factorization at small
transverse momentum can be established. The processes, therefore,
also need to be characterized by a large momentum scale, and there
has to be additionally a small measured transverse momentum.
Rigorous theoretical analyses of such reactions started from
Collins and Soper's seminal paper \cite{ColSop81}, in which they
proved factorization for di-hadron semi-inclusive processes in
$e^+e^-$ annihilation. Non-perturbative TMD fragmentation
functions were defined and then further studied along with TMD
parton distributions in \cite{ColSop81p}. The approach was
extended to Drell-Yan dimuon production at hadron colliders
\cite{ColSopSte85}.

More recently, these factorization theorems and the one for
semi-inclusive DIS have been carefully (re-)examined in the
context of the gauge-invariant definitions of the TMD parton
distributions and fragmentation functions
\cite{JiMaYu04,ColMet04}, paying in particular attention to the
``naively time-reversal-odd'' functions. In summary, QCD
factorization has been established for three classes of
semi-inclusive processes: di-hadron production in $e^+e^-$
annihilation, semi-inclusive DIS, and the Drell-Yan process. It
still remains to be seen whether factorization holds for more
complicated processes in hadronic scattering, such as for di-jet
(di-hadron) azimuthal angular correlations
\cite{BoeVog03,BomMulPij05}. These processes, too, are
characterized by a large scale (the individual jet or hadron
transverse momenta), and by an additional small transverse
momentum related, for example, to the {\it pair} transverse
momentum, or to the deviation of the two jets from being
``back-to-back''. Note that this is in contrast to {\it
single-inclusive} processes at hadron colliders like
$p_{\uparrow}p\to \pi X$. The spin asymmetries for such reactions
are power-suppressed (``higher twist''), and the theoretical
description should be based on the  methods developed in
\cite{qiu}, where factorization in terms of higher-twist
correlation functions was established.

In this paper, we will use the factorization approaches at small
transverse momentum discussed above to study the single spin
asymmetries in semi-inclusive DIS. We will focus on the Sivers and
Collins asymmetries which presently are the most interesting ones.
We realize that at the current stage it is difficult to apply the
full factorization formulas developed in the literature in fitting
the data and making predictions. Instead, we will make some
approximations, neglecting higher order terms in the hard and soft
factors. In this way we of course introduce some theoretical
shortcoming, which we hope can be overcome in future studies. Our
purpose is to look for the ``main effects'', that is, to provide a
quantitative description of the spin effects now studied
experimentally, and to draw our conclusions on the Collins and
Sivers functions from these. Another goal of this paper is to use
the information gathered from semi-inclusive DIS to make
predictions for processes at RHIC, which is now taking data in
transversely polarized $pp$ scattering. With the fitted
parameterizations for the Sivers functions that describe the
HERMES data very well, we will calculate the Sivers asymmetries
for processes at RHIC, including Drell-Yan dimuon production and
di-jet and jet-plus-photon correlations. We will demonstrate that
these asymmetries are expected to be large at RHIC and should
therefore be closely investigated in the future polarized $pp$
runs. This would then provide further tests of the physical
picture behind the SSA, and of our theoretical understanding.

To predict the SSAs at RHIC from the distributions fitted in DIS
relies on the factorization for the $pp$ processes which, as
discussed above, is so far only established for the Drell-Yan
reaction. It also relies on the universality of the TMD
distributions. This issue has been addressed in detail in
\cite{Col02,BelJiYua02,BoeMulPij03,ColMet04,BomMulPij05}. It was
found, for example, that the Sivers functions for the Drell-Yan
process will have an opposite sign compared to those for SIDIS,
as a result of the behavior of the gauge-links in the
functions under the time-reversal operation. We will use this
additional sign in our prediction for the Drell-Yan SSA at RHIC
based on our Sivers function fitted to the DIS data. However,
recent work has shown \cite{BomMulPij05} that the issue of
universality appears to be much more complicated for the case of
di-jet correlations, where the more involved color structure has
profound consequences on the gauge links. As a result, the Sivers
functions for this reaction will differ from those in DIS by more
than just a sign. This reservation notwithstanding, in order to obtain an
order of magnitude estimate we will assume in this paper that the
Sivers functions to be used for di-jet correlations have the same
sizes as those for the Drell-Yan processes, and opposite signs
with respect to the DIS Sivers functions.

In all calculations of cross sections and asymmetries below, we
will use the GRV LO parameterizations for the unpolarized quark
distributions \cite{GRV94}, and the Kretzer set of unpolarized
quark fragmentation functions \cite{Kretzer}. These will also
serve as starting points for our parameterizations of the Sivers
and Collins functions.

The rest of the paper is organized as follows. In Sec.~II, we will
review the basic formulas for the SSAs in SIDIS, and make model
parameterizations for the Sivers and Collins functions. We then
fit our parameterizations to the HERMES data. We will also compare
our fit with the recent COMPASS data on the Sivers and Collins
asymmetries. In Sec.~III, we will calculate the Sivers asymmetries
for the Drell-Yan process and for di-jet and jet-photon correlations at RHIC,
using the fitted parameterizations from Sec.~II. We summarize in
Sec.~IV.

\section{SSA in Semi-inclusive Deep Inelastic Scattering}

In this section, we will study the SSA in the SIDIS processes
$e p\to e hX$ and $\mu d\to \mu hX$, where 
$h$ represents a hadron observed in the final state. We will compare
the theoretical calculations of the asymmetries with the HERMES
measurements. We will use some simple parameterizations for the
Sivers functions and the Collins fragmentation functions, and fit
these to the experimental data. A comparison of our fit with the
COMPASS measurements will also be presented. Similar
phenomenological studies of these asymmetries have also been
performed in \cite{Metz05,Ans05} for the Sivers case and in
\cite{schweitzer} for the Collins asymmetry, using the earlier
HERMES data.

We will start by briefly recalling the factorization formulas for
the SIDIS process. For details, we refer the reader to
Ref.~\cite{JiMaYu04}. As discussed in the introduction, we will
make some simplifying approximations, in order to sharpen the
constraints on the Sivers and Collins functions.

\subsection{Theoretical Formalism and Approximations}

The differential cross section for SIDIS, including the
unpolarized part and the Sivers and Collins asymmetry
contributions, may be written in the following form:
\begin{eqnarray}
    \frac{d\sigma}{dx_Bdydz_hd^2\vec {P}_{h\perp}}
      &=& \frac{4\pi\alpha^2_{\rm em}s}{Q^4}\left[(1-y+y^2/2)
      x_B \left(F_{UU}-\sin(\phi_h-\phi_S) |\vec{S}_\perp|
      F_{UT}^{sivers}\right)\right.\nonumber\\
      &&\left.
      -(1-y)x_B|\vec{S}_\perp|\sin(\phi_h+\phi_S)
F_{UT}^{collins}\right] \ ,
\end{eqnarray}
where $\phi_h$ ($\phi_S$) is the angle between the lepton plane
and the $\gamma^*$-hadron-plane (and the transverse target spin),
$y$ is the fraction of the incident lepton energy carried by the
photon, and $\vec {P}_{h\perp}$ is the (measured) transverse
momentum of the hadron. In order to compare with the experimental
data, in the above formula and the following calculations, the
azimuthal angles ($\phi_S$ and $\phi_h$) are defined in the
so-call virtual photon frame where the virtual photon is moving in
the $z$ direction. These definitions are different from those in
\cite{JiMaYu04} where a hadron frame has been chosen to define
these angles. This difference has led to the different signs in
the above formula, compared to that in \cite{JiMaYu04}. The
structure functions $F_{UU}$ and $F_{UT}$ will depend in general
on $\vec {P}_{h\perp}$, and on the invariant mass $Q^2$ of the
virtual photon, the Bjorken variable $x_B$, and on the fraction
$z_h$ of the photon longitudinal momentum carried by the hadron
observed in the final state. According to the factorization
formula of \cite{JiMaYu04}, the structure functions can be
factorized into the TMD parton distributions and fragmentation
functions, and soft and hard factors. For example, for the
unpolarized structure function, we will have \cite{JiMaYu04}
\begin{eqnarray}
\label{crsecfac}
F_{UU}(x_B,z_h,Q^2,P_{h\perp})&=&\sum_{q=u,d,s,...} e_q^2\int
d^2\vec{k}_{\perp} d^2\vec {p}_{\perp}
      d^2\vec{\lambda}_\perp
   \nonumber \\
   && \times  q\left({x_B}, k_{\perp},\mu^2,x_B\zeta, \rho\right)
    \hat q\left({z_h}, p_{\perp},\mu^2,\hat\zeta/z_h, \rho\right)
    S(\vec{\lambda}_\perp,\mu^2,\rho) \nonumber \\
&& \times H\left(Q^2/\mu^2,\rho\right)
\delta^{(2)}(z_h\vec{k}_{\perp}+\vec{p}_{\perp} +\vec{\lambda}_\perp-
\vec{P}_{h\perp}) \ .
\end{eqnarray}
This form is valid at low transverse momentum $P_{h\perp}\ll Q$ and 
is accurate at the leading power of $P_{h\perp}^2/Q^2$. As seen 
from the $\delta$-function expressing transverse-momentum conservation,
the observed hadron's transverse momentum is generated by three
contributions: the transverse momentum $\vec{k}_\perp$
of partons in the nucleon, (described by the TMD distribution $q$), 
the transverse momentum $\vec{p}_\perp$ acquired in the fragmentation 
process (as expressed by the TMD fragmentation function $\hat{q}$), 
and the combined transverse momenta $\vec{\lambda}_\perp$ of (large-angle) 
soft-gluon radiation, embodied in the soft factor $S$. 
Each of these transverse momenta is integrated
in Eq.~(\ref{crsecfac}), but leaves its imprint in the distribution
in $P_{h\perp}$ of the observed hadron. In contrast to the 
TMD functions in~(\ref{crsecfac}), $H$ is a hard factor that
depends solely on the large scale $Q$. Furthermore, $\mu\sim Q$ is a
renormalization scale, $\rho$ a gluon rapidity cut-off parameter,
and $\zeta$ is defined as $\zeta=(2 P\cdot v)^2/v^2$, with $P$ the
target hadron momentum, taken in the ``plus''-light-cone direction,
$P=(P^+,P^-=0,\vec{P}_\perp=\vec{0}_\perp)$, and $v$ a time-like vector 
conjugate to $P$, i.e., with only a ``minus''-light-cone component. For
details, see~\cite{JiMaYu04}, where also the related definition of
$\hat\zeta$ is given. Similar factorization formulas as~(\ref{crsecfac})
can be written down for the single-transversely polarized structure functions
$F_{UT}^{sivers}$ and $F_{UT}^{collins}$. 

For simplicity, in the following numerical
calculations, we will use the leading order expressions for the
hard scattering and the soft factors, for which we have $S=H=1$.
At this order, we may also neglect the $\zeta,\hat\zeta$
dependences in the parton distributions and fragmentation
functions. All this brings us to the parton model picture for
semi-inclusive DIS \cite{MulTan96}. However, we stress that higher
order effects can be systematically and consistently studied only
within the complete factorization framework. With the above
approximations, the structure functions can be simplified
to~\cite{JiMaYu04}
\begin{eqnarray}
F_{UU}&=&\int d^2\vec{k}_{\perp} d^2\vec {p}_{\perp}q\left({x_B},
k_{\perp}\right)\hat
q\left({z_h},p_{\perp}\right)\delta^{(2)}(z_h\vec{k}_{\perp}+\vec{p}_{\perp}
- \vec{P}_{h\perp})
     \ , \nonumber\\
F_{UT}^{sivers}&=&\int d^2\vec{k}_{\perp} d^2\vec {p}_{\perp}
\frac{\vec{k}_\perp\cdot \hat{\vec{P}}_{h\perp}}{M}
q_T\left({x_B}, k_{\perp}\right)\hat q\left({z_h},p_{\perp}\right)
    \delta^{(2)}(z_h\vec{k}_{\perp}+\vec{p}_{\perp}
- \vec{P}_{h\perp})\ ,\nonumber\\
F_{UT}^{collins}&=&\int d^2\vec{k}_{\perp} d^2\vec {p}_{\perp}
\frac{\vec{p}_\perp\cdot \hat{\vec{P}}_{h\perp}}{M_h} \delta
q_T\left({x_B}, k_{\perp}\right)\delta \hat
q\left({z_h},p_{\perp}\right)
    \delta^{(2)}(z_h\vec{k}_{\perp}+\vec{p}_{\perp}
- \vec{P}_{h\perp})
 \ ,
\end{eqnarray}
where a sum over all quark and antiquark flavors, weighted with
the squared quark electric charge, is implicitly understood from
now on. $\hat{\vec{P}}_{h\perp}$ denotes a unit vector in
direction of $\vec{P}_{h\perp}$. In the above equations, $q$ and
$\hat q$ represent the unpolarized quark distribution and
fragmentation functions, respectively, $q_T$ the Sivers functions,
$\delta \hat q$ the Collins functions, and $\delta q_T$ the
transversity distribution functions. The definitions of the above
distributions and fragmentation functions are consistent with the
so-called ``Trento conventions'' \cite{Trento}, while opposite in sign
with respect to that used in \cite{Ans05} for the Sivers function. To 
optimize statistics, the experimental measurements of the asymmetries are
normally presented after integrating over the modulus of the
hadron's transverse momentum $P_{h\perp}$. After integration, the
cross section can be written as
\begin{eqnarray}
\label{eq4}
    \frac{d\sigma}{dx_Bdydz_h d\phi_h}
      &=& \frac{d\sigma_{UU}}{dx_Bdydz_h}-\sin(\phi_h-\phi_S)
\frac{d\sigma_{UT}^{sivers}}{dx_Bdydz_h}
            -\sin(\phi_h+\phi_S)
\frac{d\sigma_{UT}^{collins}}{dx_Bdydz_h} \ ,
\end{eqnarray}
where the various differential cross sections will depend on
$x_B$, $z_h$ and $y$. The dependence of the cross sections on the
azimuthal angles results in the azimuthal asymmetries measured in
experiment. The unpolarized cross section is given by
\begin{eqnarray}
    \frac{d\sigma_{UU}}{dx_Bdydz_h}
      &=& \frac{4\pi\alpha^2_{\rm
em}s}{Q^4}\left(1-y+\frac{y^2}{2}\right)
      x_B q(x_B)\hat q(z_h) \ ,\label{eq5}
\end{eqnarray}
where $q(x_B)$ and $\hat q(z_h)$ are the integrated parton
distribution and fragmentation functions. Here we assume that we
can obtain the integrated parton distribution by integrating over
the transverse momentum in the corresponding TMD parton
distribution. This assumption will of course need to be modified
if higher-order corrections are considered \cite{JiMaYu04}.
Similarly, we can calculate the polarized cross sections. In these
calculations, we further assume that the final hadron's transverse
momentum is entirely related to the transverse-momentum dependence
in the Sivers and Collins functions. The transverse momentum
contributed by the other factors in the factorized
formula~(\ref{crsecfac}) will give some smearing effects which may
be viewed as ``sub-dominant''. After this approximation, we can
write down the polarized cross sections as
\begin{eqnarray} \label{eq6a}
    \frac{d\sigma_{UT}^{sivers}}{dx_Bdydz_h}
      &=& |S_\perp|\frac{4\pi\alpha^2_{\rm
em}s}{Q^4}(1-y+\frac{y^2}{2})
      x_B q_T^{(1/2)}(x_B)\hat q(z_h) \ ,\\
    \frac{d\sigma_{UT}^{collins}}{dx_Bdydz_h}
      &=& |S_\perp|\frac{4\pi\alpha^2_{\rm em}s}{Q^4}(1-y)
      x_B \delta q_T(x_B)\delta\hat q^{(1/2)}(z_h) \ ,
      \label{eq6b}
\end{eqnarray}
where $\delta q_T$ is the integrated transversity distribution
function. $q_T^{(1/2)}(x_B)$ and $\delta\hat q^{(1/2)}(z_h)$ are
defined as
\begin{eqnarray}\label{eq7}
q_T^{(1/2)}(x_B)&=&\int d^2k_\perp\frac{|\vec{k}_\perp|}{M}
q_T\left({x_B}, k_{\perp}\right) \; ,\nonumber\\
\delta\hat q^{(1/2)}(z_h)&=&\int
d^2p_\perp\frac{|\vec{p}_\perp|}{M_h} q_T\left({z_h},
p_{\perp}\right)  \;  .
\end{eqnarray}
The above formulas (\ref{eq4})-(\ref{eq7}) will be used in the
following calculations to study the experimental data for the
asymmetries as functions of $x_B$ and $z_h$. Before doing so, we
need to set up models for the Collins and Sivers functions.

We would like to add one more comment before we proceed. In the
derivation of Eq.~(\ref{eq6a}), we have omitted the transverse
momentum dependence of the fragmentation function, which we
referred to as ``sub-dominant''. This ``sub-dominant''
contribution could become important at small $z_h$ where we cannot
neglect the influence of the transverse momentum in the
fragmentation process. For example, for a typical transverse
momentum of the final state hadron of $P_{h\perp}\approx 200\sim
300$~MeV, the quark transverse momentum could be as large as
$2\sim 3$~GeV at $z_h\approx 0.1$, much bigger than the typical
value of the intrinsic quark transverse momentum for the Sivers
function, which is of order of a few hundred MeV. This means that
at small $z_h$ our approximation will break down, and the
transverse momentum in the fragmentation function will be
important. This effect will smear out the polarized cross section
and suppress the asymmetry. Indeed, when assuming a Gaussian
transverse momentum dependence for both the distribution and the
fragmentation functions, an additional factor of $z_h$ appears,
suppressing the polarized cross section at small $z_h$
\cite{Ans05}. As a consequence, we should be cautious to apply
Eq.~(\ref{eq6a}) at small $z_h$. On the other hand, the above
drawback does not apply to the case of the polarized cross section
for the Collins contribution, Eq.~(\ref{eq6b}), where we omit the
transverse momentum dependence in the distribution. This is
because there is no kinematic enhancement associated with the
intrinsic transverse momentum in the parton distribution, compared
to the fragmentation case.

\subsection{Model for the Sivers functions}

There exist by now quite a few model calculations for the quark
Sivers functions in the nucleon \cite{siversmodel}. The results of
these vary rather widely. Here, we will instead adopt simple
parameterizations for the Sivers functions and fit these to the
HERMES data. We choose a form that has only a single free
parameter for each flavor; the present data probably do not yet
warrant a more complex form. Our parameterization is as follows:
\begin{eqnarray}
\frac{u_T^{(1/2)}(x)}{u(x)}&=&S_u x(1-x) \ ,\nonumber\\
\frac{d_T^{(1/2)}(x)}{u(x)}&=&S_d x(1-x)\ ,\label{eq9}
\end{eqnarray}
where in both equations $u(x)$ is the unpolarized $u$-quark
distribution. We assume that only the quark Sivers functions are
non-zero, and that the antiquark ones vanish. This assumption will
of course likely need to be modified at small $x$. In the above
parameterization, the factor $x$ on the right-hand-side represents
the valence nature of the Sivers function, whereas the factor
$(1-x)$ denotes an expected suppressed behavior of the function at
large $x$~\footnote{This power suppression in $(1-x)$ could
actually be as strong as $(1-x)^2$~\cite{Metz05}. In our fit, most
data are in the intermediate range of $x$, and the $x\to 1$ limit
is not really reached. In any case, the power of $(1-x)$ will be
modified by logarithms.}

\subsection{Model for the Collins functions}

For the Collins asymmetry, we have two sets of unknown functions:
the transversity distributions and the Collins fragmentation
functions. From the present experimental data on the Collins
asymmetries we cannot obtain constraints on both of them
simultaneously~\footnote{Note, however, that independent
information on the Collins functions is now coming from
measurements of hadron-pair production by the BELLE
collaboration~\cite{belle}. It is hoped that combination of these
results with those from lepton scattering would eventually give
information on transversity.}. Here we adopt a parameterization
for the transversity function \cite{Vogelsangh1} that is based on
saturation of the Soffer inequality \cite{Sof94}. We note that
this parameterization represents an upper bound for the
transversity functions for the quarks.

As for the Sivers functions, there have also been several model
calculations for the Collins functions \cite{collinsmodel},
showing rather wide variations. Again, we will just use a simple
parameterization for the $z_h$ dependence of $\delta \hat
q^{(1/2)}$. The flavor dependence of the Collins functions is
important since one would like to describe the asymmetries for
different hadron species. From the theory side, one could get
constraints for the flavor dependence based on momentum
conservation in the fragmentation process: the Sch\"afer-Teryaev
sum rules \cite{SchTer99}. These sum rules state that the
odd-moment (for the intrinsic transverse momentum) of the Collins
function vanishes when the function is summed over all hadron states.
Because the sum rules only involve the integrals over $z_h$ and $p_\perp$
(with a weight $p_\perp$) of the Collins functions,
one cannot obtain from them more detailed
constraints, e.g., for the $z_h$ or $p_\perp$ dependences. In
the following we will motivate a simple conjecture
for the Collins functions, based on quark-hadron duality in the
fragmentation process. This will provide us with additional
constraints. The main result is that {\it any quark Collins
fragmentation function is very small when summed over all final
hadrons}. For example, the $u$ quark Collins functions to all
hadron final states will satisfy
\begin{equation}
\sum_h\delta \hat u^h(z_h,p_\perp)={\cal O}(m_u)\ .
\end{equation}
The above equation is motivated as follows. The Collins
function is defined as
\begin{eqnarray}
  \frac{\epsilon^{ij}p_\perp^j}{M_h}\delta \hat q^{h}(z_h,p_\perp)
&=&\frac{n^-}{3z}
  \int \frac{d\xi^+}{2\pi}\frac{d^2 \vec{b}}{(2\pi)^2}
  e^{-i(k^-\xi^+-\vec{k}_\perp\cdot\vec{b}_\perp)}\sum_X
  \Tr\left\{\gamma_5\gamma^-\gamma^i\right.
 \label{ffdef} \\
  && \left.\times \langle 0|{\cal L}(-\infty,0)\psi(0)
   |P_hX\rangle\langle P_hX|\overline{\psi}(\xi^+,\vec{b}) {\cal
L}^\dagger(\xi^+,\vec{b},-\infty)|0\rangle\right\} \nonumber \ ,
\end{eqnarray}
where $k^-=P^-_h/z_h$ and $\vec{k}_\perp = -\vec{p}_{\perp}/z_h$.
Here, the final state hadron has been taken to have a large 
light-cone ``minus'' momentum component $P_h^-$ [we remind the reader that 
$P_h$ denotes the momentum of the observed hadron, while $\vec{p}_{\perp}$
is the (integrated) transverse momentum that the hadron acquires 
in the fragmentation process relative to the fragementing parton;
see Eq.~(\ref{crsecfac})]. Finally, 
${\cal L}$ is the gauge link along the light-cone direction
conjugate to $P_h$. Here,  Now, if we sum over all
hadrons in the above Collins fragmentation functions, the
intermediate hadronic states can be replaced by a quark or a quark
plus gluons (or quark-antiquark pairs) using quark-hadron duality
arguments. We then get the following equation:
\begin{eqnarray} \label{cdual}
  \sum_h\frac{\epsilon^{ij}p_\perp^j}{M_h}\delta \hat
q^{h}(z_h,p_\perp) &=&\frac{n^-}{3z}
  \int \frac{d\xi^+}{2\pi}\frac{d^2 \vec{b}}{(2\pi)^2}
  e^{-i(k^-\xi^+-\vec{k}_\perp\cdot\vec{b}_\perp)}\sum_X
  \Tr\left\{\gamma_5\gamma^-\gamma^i\right.
  \\
  && \left.\times \langle 0|{\cal L}(-\infty,0)\psi(0)
   |P_qX\rangle\langle P_qX|\overline{\psi}(\xi^+,\vec{b}) {\cal
L}^\dagger(\xi^+,\vec{b},-\infty)|0\rangle\right\} \nonumber \ ,
\end{eqnarray}
whose validity rests on the argument of quark-hadron duality for
the fragmentation process. Duality-breaking effects will somewhat
modify the above equation. The right hand side of
Eq.~(\ref{cdual}) may be viewed as a quark Collins fragmentation
function into a quark (or antiquark/gluon) state. The
helicity-flip required for a non-vanishing Collins function is
then possible because of a finite quark mass. Thus, we
approximately expect
\begin{equation}
\sum_h\delta \hat q^h(z_h,p_\perp)={\cal O}(m_q)\approx 0\ .
\label{collinsconj}
\end{equation}
If we further assume that the fragmentation functions for $u$ and
$d$ quarks to strange mesons are suppressed relative to those into
pions, we can have even stronger constraints for the pion Collins
functions:
\begin{equation}
\delta \hat q^{\pi^+}(z_h,p_\perp)+\delta \hat
q^{\pi^-}(z_h,p_\perp)+\delta \hat q^{\pi^0}(z_h,p_\perp)\approx
0\ , \label{collinsconstr}
\end{equation}
where $q$ represents any flavor of $u$,$d$ quarks and their
antiquarks. Further simplification of the above equation can be
derived by considering isospin and charge symmetry relations
between the different fragmentation functions. For example, we
will have the following relations,
\begin{eqnarray}
&&\delta \hat u^{\pi^+}=\delta \hat d^{\pi^-}=\delta \hat{\bar
d}^{\pi^+}=\delta
\hat {\bar u}^{\pi^-}\equiv\delta \hat q^\pi_{fav.}\nonumber\\
&&\delta \hat d^{\pi^+}=\delta \hat u^{\pi^-}=\delta \hat {\bar
u}^{\pi^+}=\delta\hat
{\bar d}^{\pi^-}\equiv\delta \hat q^\pi_{unfav.}\nonumber\\
&&\delta \hat u^{\pi^0}=\delta \hat d^{\pi^0}=\delta \hat {\bar
d}^{\pi^0}=\delta \hat {\bar u}^{\pi^0}=\frac{1}{2}\left[\delta
\hat q^\pi_{fav.}+\delta \hat q^\pi_{unfav.}\right] \ .
\end{eqnarray}
Here $\delta \hat q^\pi_{fav.}$ and $\delta \hat q^\pi_{unfav.}$
represent the ``favored'' (in the sense that the leading Fock
state of the hadron contains the parent quark flavor) and
``unfavored'' (where it does not contain it) fragmentation
functions, respectively. Substituting the above relations into
Eq.~(\ref{collinsconstr}), we will obtain
\begin{equation}
\delta \hat q^\pi_{fav.}+\delta \hat q^\pi_{unfav.}\approx 0\ ,
\label{ucollins}
\end{equation}
which means that the unfavored Collins function is approximately
equal to the favored one with opposite sign. This result is
counter to the usual notion in the literature that the favored
fragmentation functions should be much larger than the unfavored
ones. It of course crucially depends on the validity of the
approximations made in the above derivation and will be subject to
some corrections. We note, however, that this observation has also
support from a string model description for the Collins
fragmentation function \cite{Artru}. As a test, we will treat in
the following the favored and unfavored Collins functions free
from the constraint (\ref{ucollins}), and fit them to the data to
see if they naturally satisfy the above relation or not. To
parameterize the Collins functions, we use the following two sets
of functional forms,
\begin{eqnarray}
{\rm Set~I:~~~}\delta \hat q_{fav.}^{\pi(1/2)}(z)&=&C_f z(1-z)
\hat u^{\pi^+}(z)\ ,~~~ \delta \hat q^{\pi(1/2)}_{unfav.}(z)=C_u
z(1-z)\hat u^{\pi^+}(z) \
, \nonumber \\
{\rm Set~II:~~~}\delta \hat q_{fav.}^{\pi(1/2)}(z)&=&C_f z(1-z)
\hat u^{\pi^+}(z)\ ,~~~ \delta \hat q^{\pi(1/2)}_{unfav.}(z)=C_u
z(1-z)\hat d^{\pi^+}(z) \ . \label{collinspar}
\end{eqnarray}
The $z$ factor in these parameterizations represents the vanishing
of the Collins function at small $z$, and the $(1-z)$ factor
follows arguments made in \cite{Col93}. The difference between
these two sets is that for Set I we parameterize both favored and
unfavored Collins functions in terms of the favored unpolarized
quark fragmentation function, while for Set II we parameterize the
unfavored Collins function using the unfavored unpolarized quark
fragmentation function. Set I is inspired by the constraint of
Eq.~(\ref{ucollins}); note that this ansatz is expected to violate the
positivity constraints at very large $z_h$. On the other hand, Set
II respects the positivity constraints, provided $|C_{f,u}|\leq 4$. 
In the following, we will fit the HERMES data with these two sets
of parameterizations.

\subsection{Comparison with SIDIS Data}

We will now calculate the Collins and Sivers asymmetries using the
functions specified above, and fit the free parameters to the new
HERMES data on the Collins and Sivers  asymmetries
\cite{HERMESnew}. Here we will use $Q^2=2.41$~GeV$^2$, which is
the average for the HERMES kinematics. We choose $\mu=Q$ in the
unpolarized parton distribution and fragmentation functions. We
will then compare our fit results to the recent measurements by
COMPASS~\cite{COMPASS}. For the fitting, we use the CERNLIB MINUIT
routine~\cite{cernlib}.

As a function of $x_B$, the Sivers asymmetry can be calculated
from the following formula:
\begin{eqnarray}
A_N(x_B)=-\frac{\int dz_hd
y\frac{d\sigma_{UT}^{sivers}}{dx_Bdydz_h}}{\int dz_hd
y\frac{d\sigma_{UU}}{dx_Bdydz_h}} \ ,
\end{eqnarray}
where the minus sign results from the minus sign in the polarized
differential cross section in Eq.~(\ref{eq4}). Since the $y$
integral is the same for the numerator and denominator, it cancels
out. Moreover, the integral over $z_h$ can be factored out as a
consequence of the approximations that led to Eqs.~(\ref{eq5}) and
(\ref{eq6a}), and the Sivers asymmetry will be proportional just
to the ratio of the Sivers functions over the unpolarized quark
distributions, summed appropriately over flavors.

The Sivers asymmetry as a function of $z_h$ can be calculated
similarly.
\begin{figure}[t]
\begin{center}
\includegraphics[height=10cm]{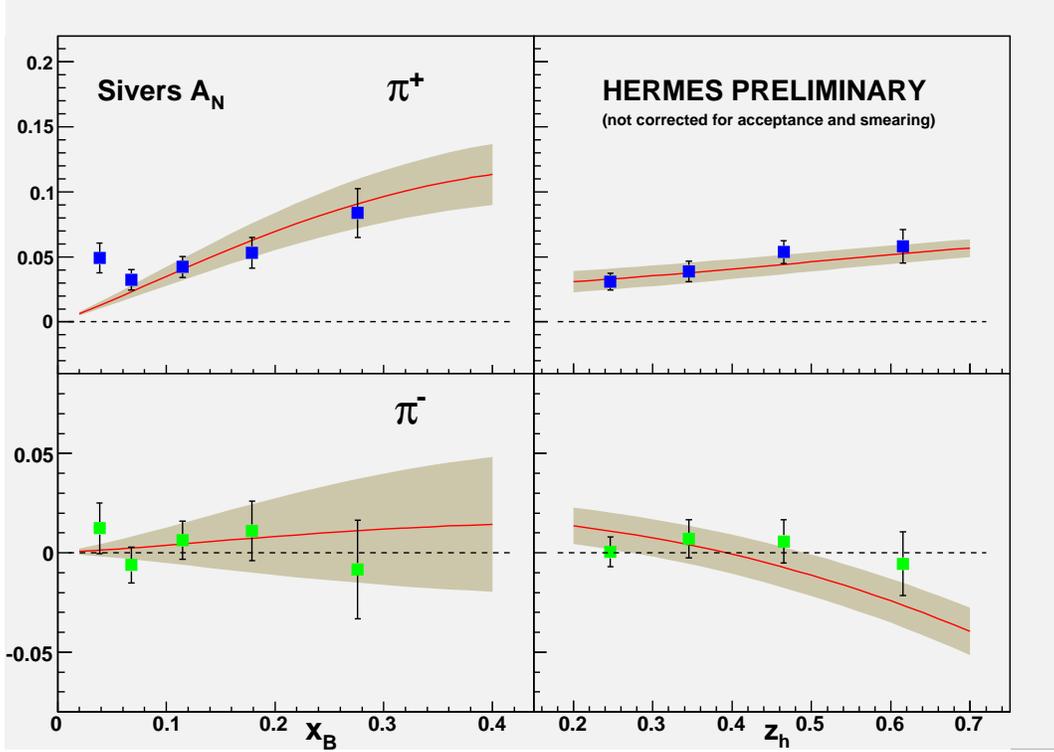}
\end{center}
\vskip -0.4cm \caption{\it Sivers SSA fit to the HERMES
data~\cite{HERMESnew}; see text. The bands correspond to the
1-$\sigma$ error of the fitted parameters. Note that the
data have not yet been corrected for acceptance and smearing. \label{fig1}}
\end{figure}
In Fig.~\ref{fig1}, we show the results of our fit of the Sivers
asymmetries for $\pi^+$ and $\pi^-$ to the HERMES data
\cite{HERMESnew}. For the two free parameters the fit gives
\begin{equation}
S_u=-0.81\pm 0.07,~~~S_d=1.86\pm 0.28 \ ,\label{eq19}
\end{equation}
with $\chi^2/{\rm d.o.f}\approx 1.2$. The band in each plot of
Fig.~1 corresponds to a 1-$\sigma$ error in the determined
parameters. One can see that the $u$ quark Sivers function appears
somewhat better constrained by the data than the $d$ quark one.
This is readily understood from the fact that $u$ quarks in DIS
enter with the charge factor $4/9$, so that in scattering off a
proton target the $u$ quark distribution is particularly selected.
Another feature is that the $d$ Sivers function comes out larger
(by about a factor of two) than the $u$ quark one, and with
opposite sign. This behavior is quite different from model
calculations \cite{siversmodel}. The result is due to the fact
that the HERMES Sivers asymmetry for $\pi^-$ is much smaller than
that for $\pi^+$. Theoretically, however, $\pi^-$ production
should also have a significant contribution from $u$ quarks,
because one finds $\hat u^{\pi^-}\approx 0.6\hat u^{\pi^+}$ for
the fragmentation functions when integrated over the
experimentally relevant region $0.2<z<0.7$. To obtain a much
smaller asymmetry for $\pi^-$ than for $\pi^+$, there then have to
be fairly strong cancellations between the $u$ and $d$ quark
Sivers functions. We note that the signs we find for our Sivers
functions are consistent with expectations in~\cite{Burkardt},
where they were qualitatively related to the opposites of the
quark contributions to the proton anomalous magnetic moment.

\begin{figure}[t]
\begin{center}
\includegraphics[height=6cm]{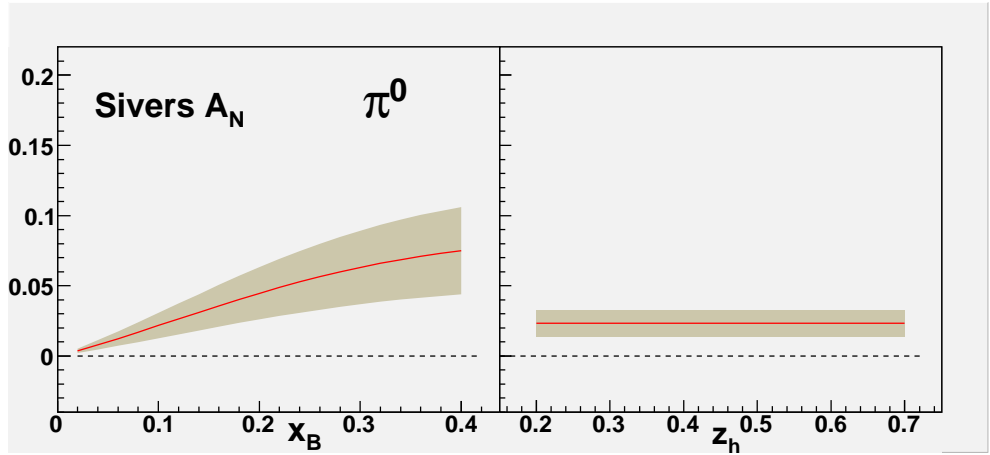}
\end{center}
\vskip -0.4cm \caption{\it Predicted Sivers SSA asymmetries for
$\pi^0$ production at HERMES. \label{fig2}}
\end{figure}

Figure~\ref{fig2} shows predictions for the $\pi^0$ Sivers
asymmetries as functions of $x_B$ and $z_h$, based on our fits for
the Sivers functions. We note that our prediction for the $\pi^0$
asymmetry is nearly independent of $z_h$. This is because the
$u$ and $d$ quark fragmentation functions for $\pi^0$ are
the same, and because in our approximation the distribution
and fragmentation functions are decoupled. However, we have to keep in mind
that this decoupling might break down at small $z_h$, as we
discussed before. This could be tested by future HERMES data.

In \cite{Metz05,Ans05}, earlier HERMES results \cite{HERMESt} for
the Sivers asymmetries were fitted. The methods somewhat differed
from ours. In \cite{Ans05}, a particular transverse momentum
dependence is assumed for the Sivers functions and the unpolarized
quark distribution and fragmentation functions. With more free
parameters the experimental data are fitted equally well, and the
$u$ and $d$ valence Sivers functions are obtained from the fit. In
\cite{Metz05}, the asymmetries weighted with the transverse
momentum of the hadron were used for the fit. Both fits find a
large $d$ quark Sivers function with opposite sign relative to the
$u$-quark one. We have also checked that our fit results for the
Sivers functions are consistent with these fits within the current
large uncertainties, where we notice that the Sivers function in
\cite{Ans05} has an opposite sign compared to ours and to the ``Trento
conventions'' \cite{Trento}.

\begin{figure}[t]
\begin{center}
\includegraphics[height=10cm]{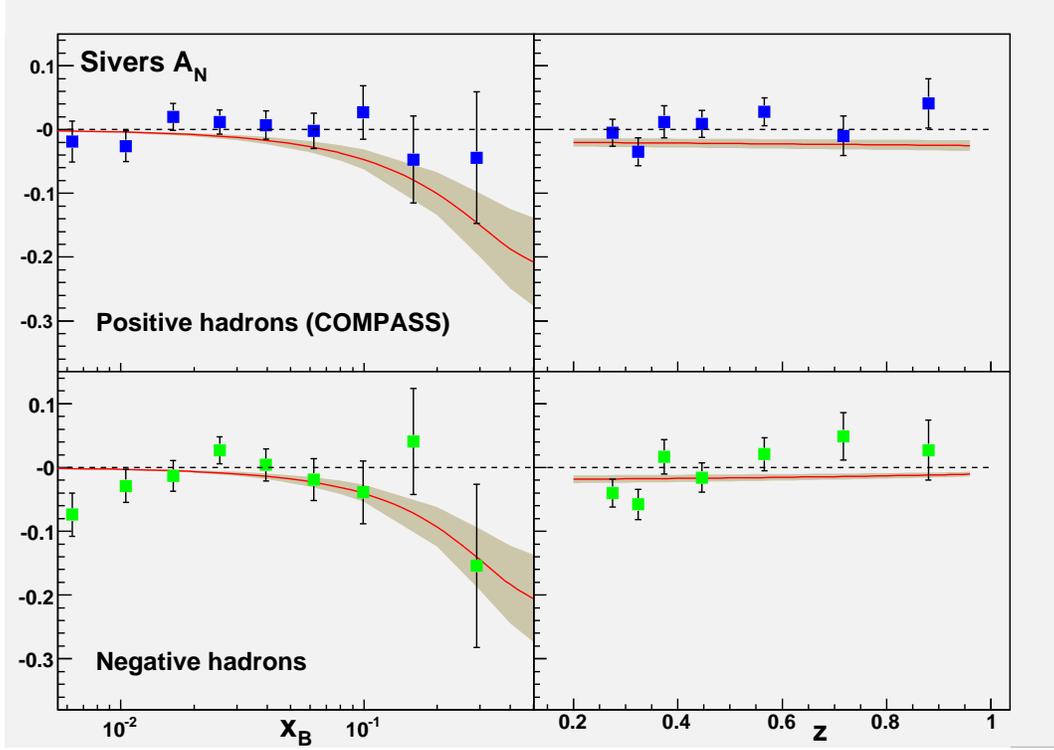}
\end{center}
\vskip -0.4cm \caption{\it Sivers asymmetries compared to the COMPASS
data~\cite{COMPASS}. \label{fig3} }
\end{figure}

The COMPASS collaboration also has measured the Sivers
asymmetry~\cite{COMPASS}, separately for positively and negatively
charged hadrons, produced off a deuteron target. To simplify the
comparison with their data, we assume that the leading hadrons are
mostly pions. We calculate the Sivers asymmetries for $\pi^+$ and
$\pi^-$ in the kinematic region of the COMPASS experiment, using
the above fitted Sivers functions for $u$ and $d$ quarks, and
compare to their data for leading positive and negative hadrons,
respectively. We show this comparison in Fig.~\ref{fig3}. One can
see that our calculations based on fits to the HERMES data
are also consistent with the COMPASS data,
within error bars. We note that for the kinematical region of the
COMPASS experiment, our predicted Sivers asymmetries for a
deuteron target are very small, except in the large-$x$ valence
region. The smallness of the Sivers asymmetry is again related to
cancellations between $u$ and $d$ contributions, which for
deuterons enter in a different combination than for a proton
target. It will be very interesting to check these predictions
with future COMPASS data for a proton target. Thanks to the higher
$Q^2$, such data would also help in confirming the leading-twist
nature of the Sivers and Collins asymmetries.

We next turn to the Collins asymmetry. Here we follow a similar
procedure as we did for the Sivers case above. As we mentioned
earlier, the situation is more complicated because of the fact the
nucleon transversity densities are currently not known, and we
need to resort to a model or ansatz for the latter. As described above, 
we will use the parameterizations for the quark transversity
distributions of~\cite{Vogelsangh1}, which represent upper
bounds for the densities. We will fit to the HERMES data using the two sets
of simple parameterizations for favored and unfavored Collins
functions given in Eq.(\ref{collinspar}).

\begin{figure}[t]
\begin{center}
\includegraphics[height=10cm]{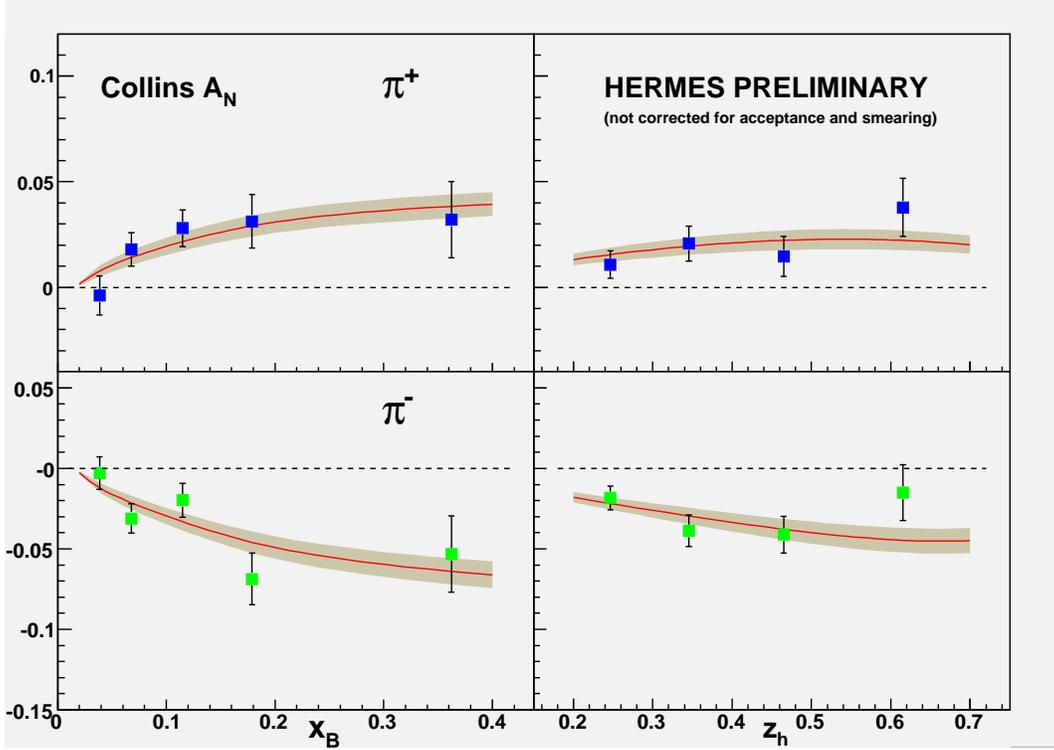}
\end{center}
\vskip -0.4cm \caption{\it Same as Fig.~\ref{fig1} for the Collins
asymmetries, using the Set I parameterization of the Collins
functions. The data~\cite{HERMESnew}
and the theory curves are for the so-called
lepton beam asymmetries. Note that the
data have not yet been corrected for acceptance and smearing.\label{fig4}}
\end{figure}

\begin{figure}[]
\begin{center}
\includegraphics[height=10cm]{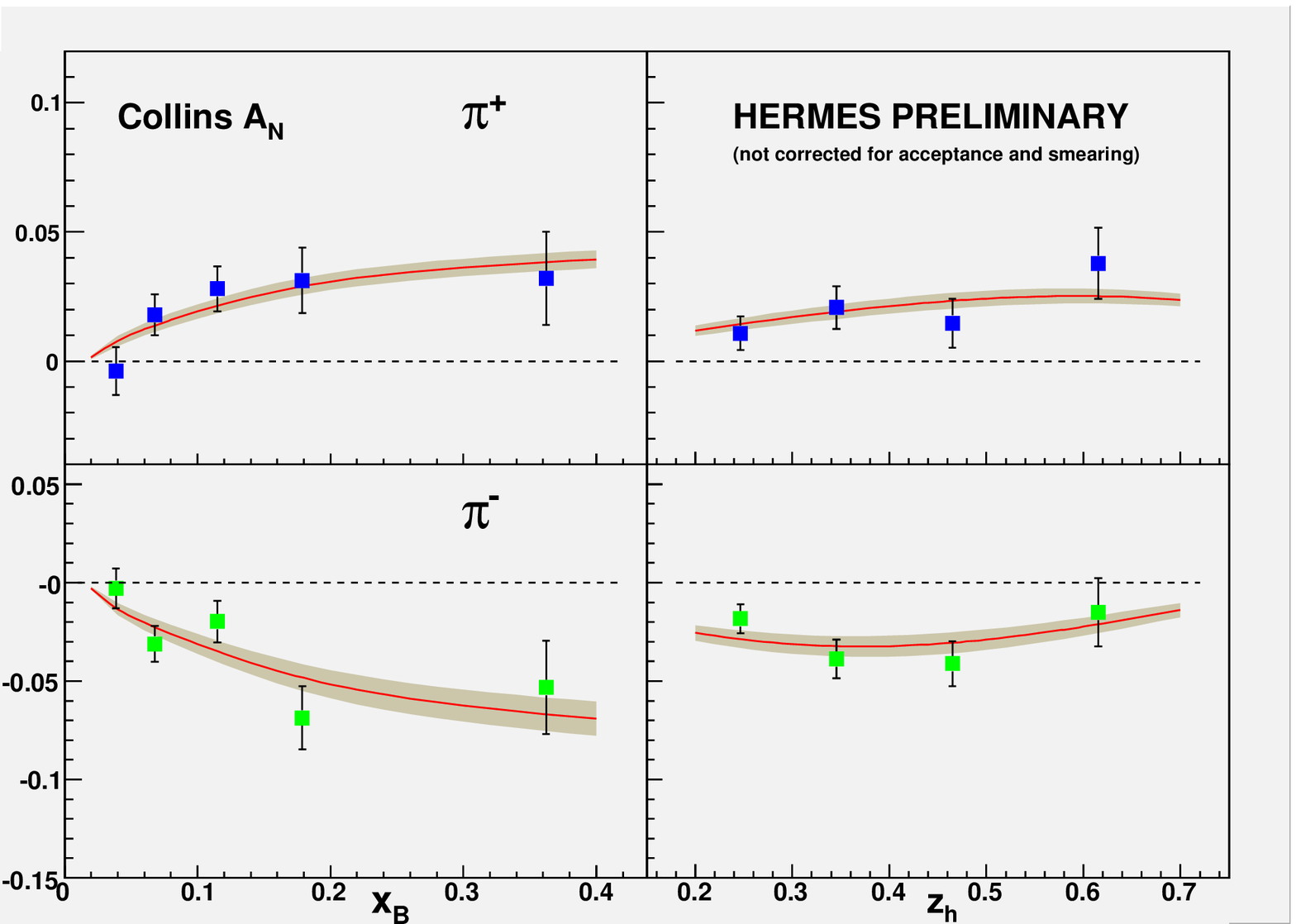}
\end{center}
\vskip -0.4cm \caption{\it Same as Fig.~\ref{fig4}, for the Set II
parameterizations for the Collins functions. \label{fig5}}
\end{figure}

The asymmetry as a function of $x_B$ is calculated from the
formula
\begin{eqnarray}
A_N^h(x_B)=-\frac{\sum\limits_{q=u,d} e_q^2 \delta \hat q^{(1/2)h}
\int dy\frac{1-y}{x_B^2y^2} x_B\delta q_T (x_B)
}{\sum\limits_{q=u,d,\bar u,\bar d} e_q^2 \hat q^h\int
dy\frac{1-y+y^2/2}{x_B^2y^2}x_Bq(x_B)} \ ,
\end{eqnarray}
where again the minus sign comes from the sign in the
polarized differential cross section, Eq.~(\ref{eq4}). $\delta \hat
q^{(1/2)h}$ and $\hat q^h$ represent the fragmentation functions
integrated over the accessed region in $z_h$. Kinematic cuts
impose a correlation between $x_B$ and $y$, and the integral over
$y$ will depend on $x_B$. In the experimental analysis, the data
for the Collins asymmetries are presented in two different ways.
One is to give results in terms of the virtual-photon asymmetry,
factoring out the term $(1-y)/(1-y+y^2/2)$. The other way is to
give the directly measured lepton-beam asymmetry. In our
calculations, we follow the latter way. We neglect the
contribution of longitudinal photons to the unpolarized cross
section, which HERMES has considered in the analysis of the
virtual-photon asymmetries \cite{HERMESnew}. In view of the
overall uncertainties, this is a minor effect, as we have checked
by comparing also to the virtual-photon asymmetries. From the fit
to the lepton-beam asymmetry data, we get the two fit parameters
as follows:
\begin{eqnarray}
{\rm Set~I:~~~}&&C_f=-0.29\pm 0.04,~~~C_u=0.33\pm 0.04 \ ,\\
{\rm Set~II:~~~}&&C_f=-0.29\pm 0.02,~~~C_u=0.56\pm 0.07 \ ,
\end{eqnarray}
with $\chi^2/{\rm d.o.f.}\approx 0.8 (0.7)$ for the Set I and Set
II parameterizations, respectively. The fit results are shown in
Figs.~\ref{fig4} and~\ref{fig5}, compared to the HERMES data. Both
fits are of the same quality.

\begin{figure}[]
\begin{center}
\includegraphics[height=6cm]{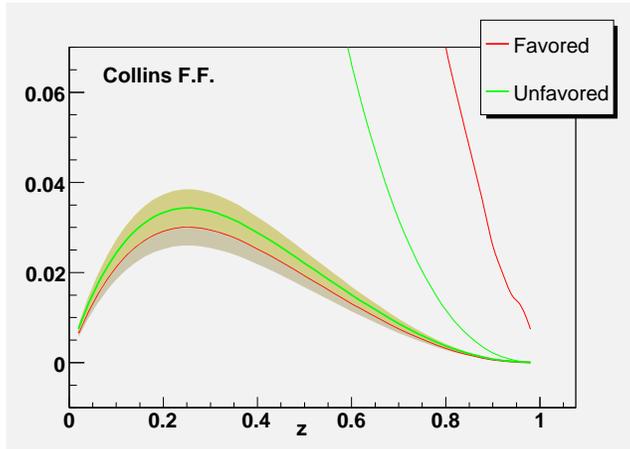}
\end{center}
\vskip -0.4cm \caption{\it Set I $\,(-1)\times$favored and unfavored
Collins fragmentation functions. Also shown are the
unpolarized quark fragmentation functions from Kretzer's
parameterizations. \label{fig9} }
\end{figure}

\begin{figure}[]
\begin{center}
\includegraphics[height=6cm]{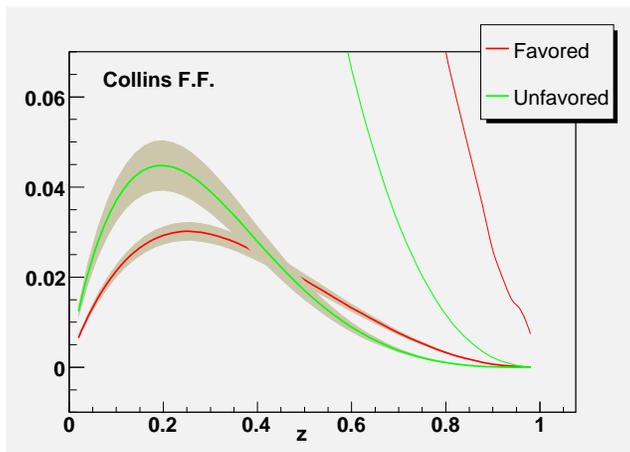}
\end{center}
\vskip -0.4cm \caption{\it Same as Fig.~\ref{fig9} for the Set II
Collins fragmentation functions. \label{fig10}}
\end{figure}

In Figs.~\ref{fig9} and~\ref{fig10}, we plot the fitted favored
and unfavored Collins functions (times $z$) for Sets I and II
respectively. Note that we multiply the favored ones by $(-1)$ to
compare their magnitudes. For comparison, we also show the
corresponding unpolarized quark fragmentation functions
\cite{Kretzer}. It is evident that the two sets of Collins
functions indeed both satisfy the positivity constraints. The
equal quality of the fits obtained for sets I and II implies that
the current experimental data neither necessarily support the
constraints we derived in Eq.~(\ref{ucollins}), nor do they rule
them out. However, from both fits we indeed find that in a quite
large range of $z_h$ the unfavored Collins function has the same
size as that of the favored one with opposite sign. A similar
conclusion was obtained from a fit to this asymmetry using the
transversity functions calculated in the Chiral Quark Model
\cite{schweitzer}. We hope that higher-statistics data will become
available in the near future that will test the relations.

\begin{figure}[t]
\begin{center}
\includegraphics[height=6cm]{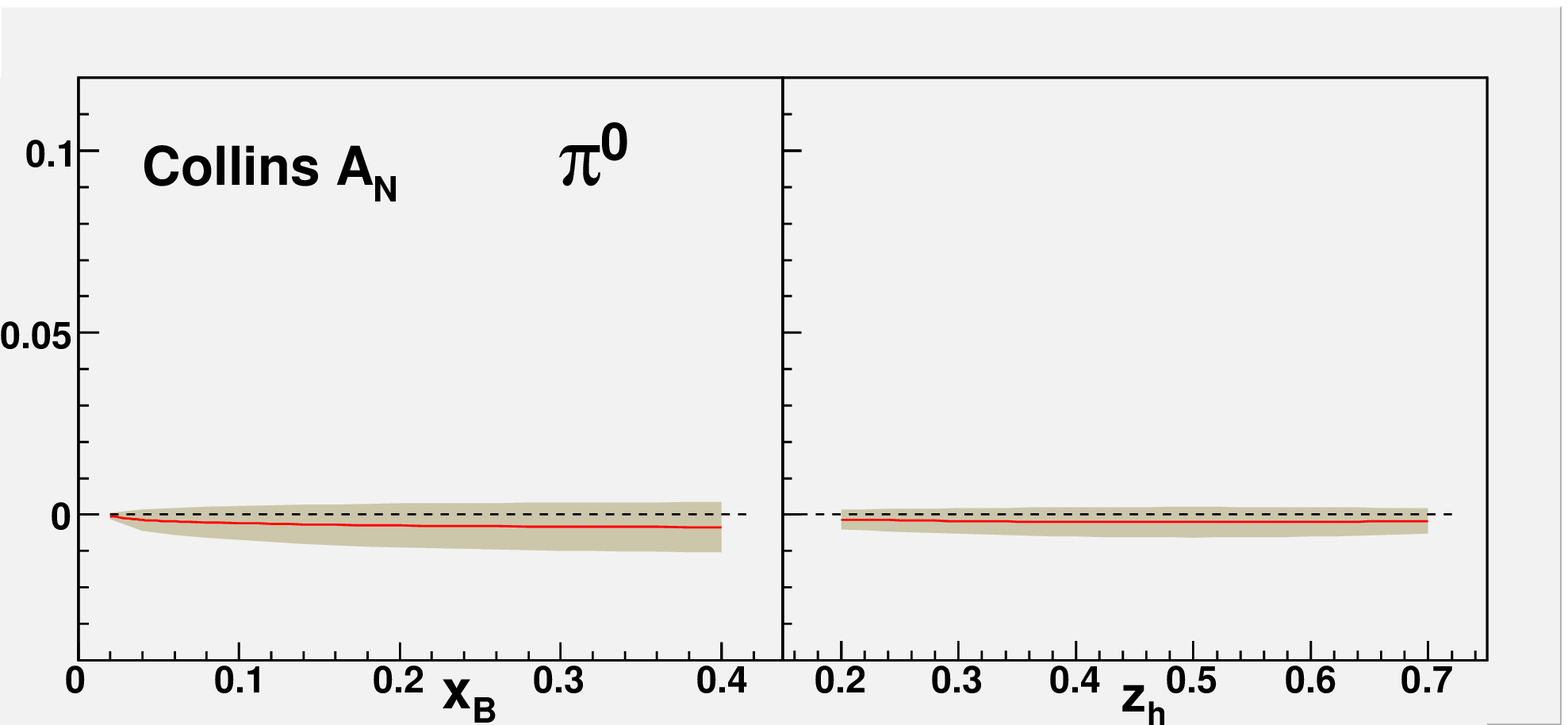}
\end{center}
\vskip -0.4cm \caption{\it Predicted Collins SSA asymmetries for
$\pi^0$ production at HERMES with Set I. \label{fig6}}
\end{figure}

\begin{figure}[t]
\begin{center}
\includegraphics[height=6cm]{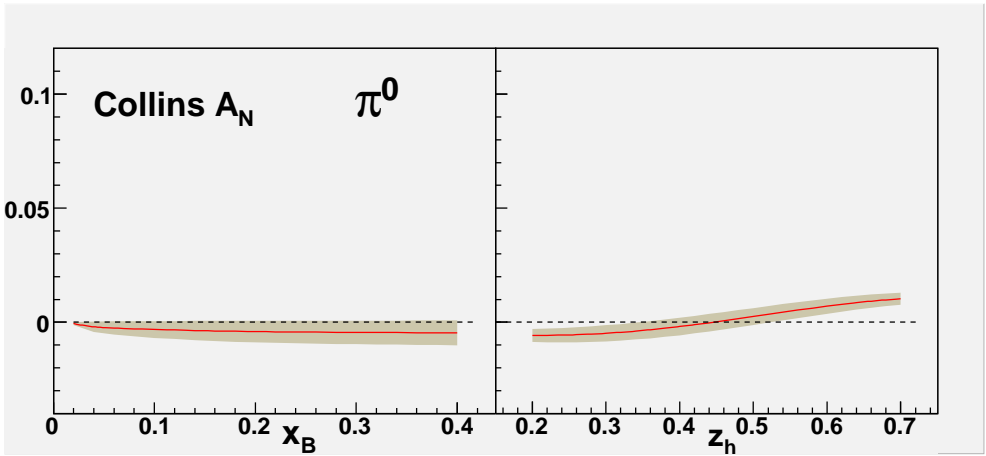}
\end{center}
\vskip -0.4cm \caption{\it Predicted Collins SSA asymmetries for
$\pi^0$ production at HERMES with Set II. \label{fig6p}}
\end{figure}

Figures~\ref{fig6} and~\ref{fig6p} show predictions for the
$\pi^0$ Collins asymmetries as functions of $x_B$ and $z_h$, based
on our fits for the Collins functions. From these plots, we find
that the asymmetries are very small for both sets of the Collins
functions, because of strong cancellations between the contributions
from favored and unfavored Collins functions.  We note that
preliminary data from HERMES indeed indicate that the
$\pi^0$ asymmetry is consistent with zero \cite{Miller}.
Higher-statistics data on this will be highly interesting for testing
the above conclusions.

\begin{figure}[]
\begin{center}
\includegraphics[height=10cm]{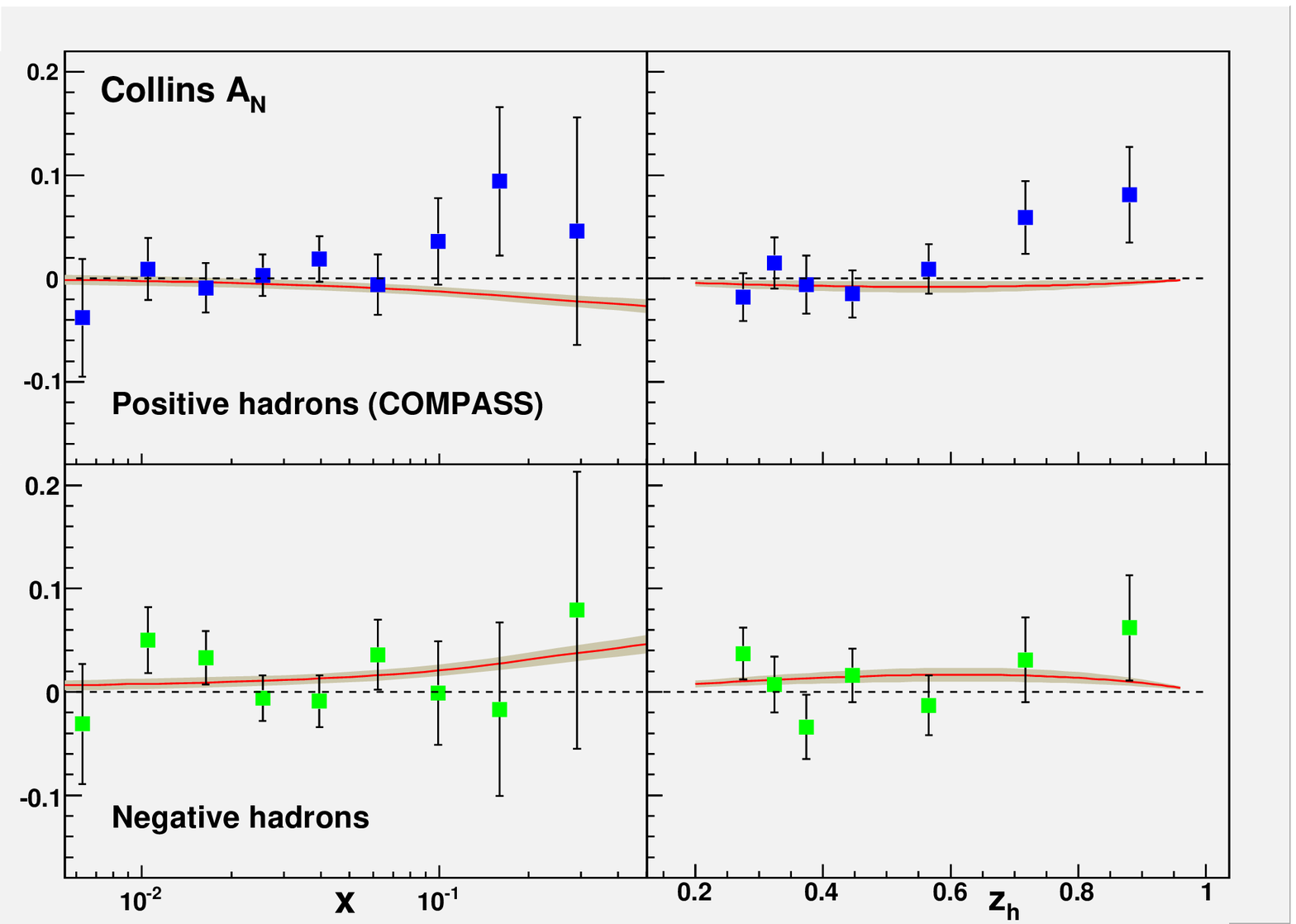}
\end{center}
\vskip -0.4cm \caption{\it Collins asymmetry compared with
COMPASS data~\cite{COMPASS} for Set I of our fitted
Collins functions. \label{fig7} }
\end{figure}

\begin{figure}[]
\begin{center}
\includegraphics[height=10cm]{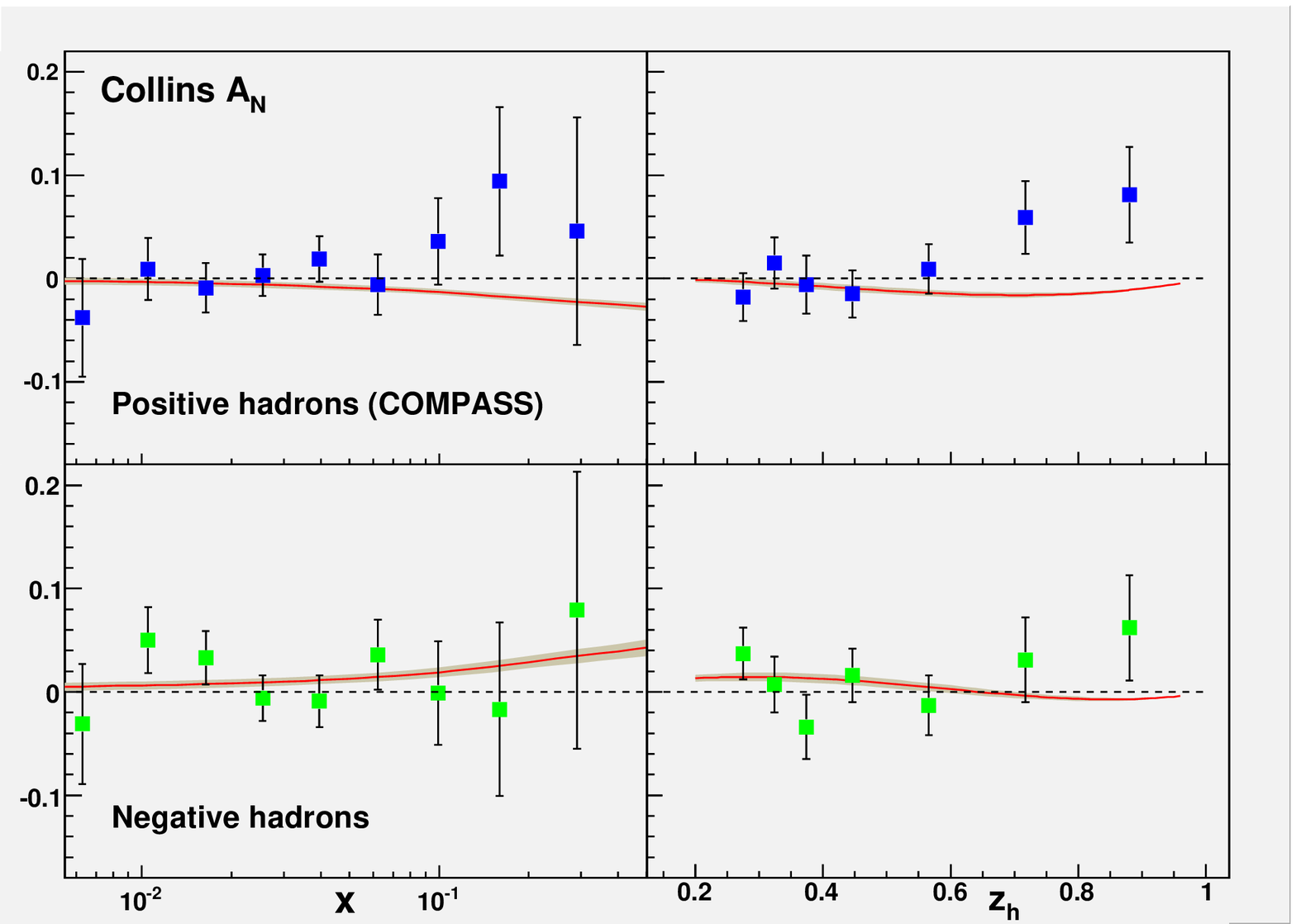}
\end{center}
\vskip -0.4cm \caption{\it Same as Fig.~\ref{fig7}, but for Set II
of Collins fragmentation functions. \label{fig8} }
\end{figure}

As before, we also compare our fit to the COMPASS
measurements~\cite{COMPASS} for the Collins asymmetries, where
only the virtual-photon asymmetries are presented. We also note
that the convention for the Collins asymmetry used by the COMPASS
Collaboration is different from that used by HERMES. 
In our calculations, we have
made the relevant modifications in order to compare with the
COMPASS data. We show these comparisons in Figs.~\ref{fig7}
and~\ref{fig8} for our Set I and II Collins functions,
respectively. As for the Sivers case above, there is good
consistency. The overall asymmetries are again small because of
the deuteron target used, and because the assumed transversity sea
quark distributions are small. Future high-statistics COMPASS data
for a proton target would be highly interesting.

\section{Single-Transverse Spin Asymmetries at Hadron Colliders}

An important issue in the study of hard-scattering processes is
the universality of the nonperturbative objects: the parton
distributions and fragmentation functions. In the case of
single-spin asymmetries, if universality holds, the Sivers
functions obtained from, for example, SIDIS can be used to predict
single-spin asymmetries in $pp$ or $\bar{p}p$ scattering. The
universality between different classes of processes is a
complicated and interesting issue that has attracted much interest
recently \cite{Metz02,BomMulPij04,ColMet04,BomMulPij05}. As we
described earlier, it was found that, while strict universality is
violated already when going from SIDIS to the Drell-Yan process,
the Sivers functions for the two processes differ only by a sign.
It is therefore possible to use the fitted Sivers function from
the last section and to predict the Sivers single-spin asymmetry
for the Drell-Yan process at hadron colliders. More complicated
processes in hadronic scattering, such as the SSA in di-jet
angular correlations \cite{BoeVog03}, are not yet completely
understood at present, as far as factorization and universality
are concerned. Progress has been made recently~\cite{BomMulPij05};
it appears that similarly to the Drell-Yan process universality is
violated only by terms that are calculable from the color
structure of the partonic scattering and hence may be taken into
account in phenomenological analyses. Below, we will also give
estimates for Sivers contributions to SSAs in di-jet correlations
and in jet-plus-photon correlations at RHIC, using the Sivers
functions of Section~II, and the usual unpolarized hard-scattering
functions. In the light of Ref.~\cite{BomMulPij05}, we expect that
our estimates will likely need to be revised once these reactions
will be completely understood in the context of factorization and
universality, to take into account the appropriate factors
embodying the nonuniversality of the Sivers functions. We shall
briefly return to this point below.

\subsection{Drell-Yan Dimuon Production $p^{\uparrow}p\to \mu^+\mu^-X$}
In this subsection, we will calculate the Sivers single-spin
asymmetry for the Drell-Yan process at RHIC,
using the fit result of the last section [see Eqs.~(\ref{eq9}) 
and~(\ref{eq19})]. As just discussed, one has
\begin{equation}
q_T^{DY}=-q_T^{DIS} \ .
\end{equation}

\begin{figure}[t]
\begin{center}
\includegraphics[height=7cm]{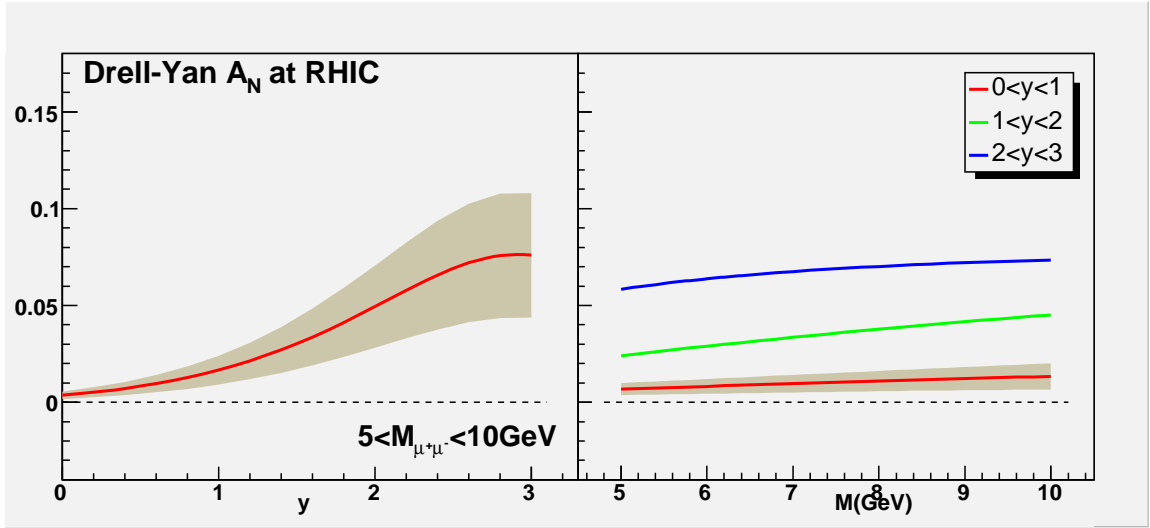}
\end{center}
\vskip -0.4cm \caption{\it Sivers asymmetries for the Drell-Yan
process at RHIC, as functions of virtual-photon rapidity $y$ and
invariant mass $M$. \label{fig11}}
\end{figure}

After integrating out the lepton angles in the rest frame of the
virtual photon, we obtain the following differential cross section
for the Drell-Yan process:
\begin{eqnarray}
\frac{d\sigma}{dM^2dyd^2q_\perp}=\frac{4\alpha^2
\pi}{3sM^2}\left[W_0(x_1,x_2,M^2,q_\perp)+\sin\phi
W_{TU}(x_1,x_2,M^2,q_\perp)\right]\ ,
\end{eqnarray}
where $M$ is the invariant mass of the lepton pair, $q_\perp$ the
virtual photon's transverse momentum, and $y$ its rapidity.
$\phi=\phi_{\gamma^*}-\phi_S$ is the difference between the
azimuthal angles of the virtual photon and the transverse
polarization vector in a frame where the polarized hadron is
moving in the $z$ direction. At low transverse momentum, $x_1$ and
$x_2$ are related to the mass and rapidity through
$x_1=M/\sqrt{s}\,e^{y}$ and $x_2=M/\sqrt{s}\, e^{-y}$ where $s$ is
the hadronic center-of-mass energy squared. According to the
factorization theorem~\cite{JiMaYu04}, the hadronic tensors $W_0$
and $W_{UT}$ can be factorized into the TMD parton distributions,
and soft and hard-scattering factors. Again, neglecting the soft
factor and using the Born expression for the hard part, we obtain
simple expressions for the tensors:
\begin{eqnarray}
W_0&=&\sum\limits_{q}\frac{e_q^2}{3}\int
d^2k_{1\perp}d^2k_{2\perp}
 q(x_1,k_{1\perp}) \overline
q(x_1,k_{2\perp})\delta^{(2)}(\vec{k}_{1\perp}+\vec{k}_{2\perp}-
\vec{q}_\perp) \ ,\nonumber\\
W_{TU}&=&\sum\limits_{q}\frac{e_q^2}{3}\int
d^2k_{1\perp}d^2k_{2\perp} \frac{\vec{k}_{1\perp}\cdot
\hat{\vec{q}}_\perp}{M} q_T(x_1,k_{1\perp}) \overline
q(x_1,k_{2\perp})\delta^{(2)}(\vec{k}_{1\perp}+\vec{k}_{2\perp}-
\vec{q}_\perp)\ ,
\end{eqnarray}
where $q_T(x_1,k_{1\perp})$ is now the Sivers function for the
Drell-Yan process. A further approximation can be made by
integrating out the transverse momentum $|\vec{q}_\perp|$, but
keeping the dependence on azimuthal angle. The differential cross
section can then be written as
\begin{equation}
\frac{d\sigma}{dM^2dyd\phi}=\frac{4\alpha^2
\pi}{3sM^2}\left[\tilde{W}_0(x_1,x_2,M^2)+\sin\phi
\tilde{W}_{TU}(x_1,x_2,M^2)\right]\ ,
\end{equation}
where
\begin{equation}
\tilde{W}_0=\sum\limits_{q}\frac{e_q^2}{3}q(x_1)\overline q(x_2) \
,
\end{equation}
and
\begin{equation}
\tilde{W}_{UT}=\sum\limits_{q}\frac{e_q^2}{3}q_T^{(1/2)}(x_1)\overline
q(x_2) \ .
\end{equation}
In Fig.~\ref{fig11}, we plot the $\sin\phi$ asymmetries as
functions of the photon rapidity $y$ and the invariant mass $M$.
From this plot, we see that the Sivers SSA asymmetry for the
Drell-Yan process at RHIC is expected to be sizable for large
rapidity, and should be measurable at RHIC if enough statistics
can be accumulated in transverse-spin running. We note that the
Sivers asymmetry in the Drell-Yan process is also a particular
focus for proposed measurements in polarized $\bar{p}p$ scattering
at the planned GSI-FAIR
facility~\cite{Metz05,PAX,Bianconi:2005bd}.

\subsection{Correlations in $p^{\uparrow}p\to {\rm jet}_1
(\vec{P}_{1\perp})+{\rm jet}_2(\vec{P}_{2\perp})+X$}

Other interesting observables at hadron colliders from which one
can access the intrinsic transverse-momentum dependence of parton
distributions are ``back-to-back'' correlations between two
jets~\cite{BoeVog03,BomMulPij05}. More specifically, we are
interested in situations in which the sum of the two jet
transverse momenta,
$\vec{q}_\perp\equiv\vec{P}_{1\perp}+\vec{P}_{2\perp}$ (or a
component or projection thereof), is measured, while both
$P_{1\perp}$ and $P_{2\perp}$ individually are large.
As for the Drell-Yan process discussed above, this is usually a
small transverse momentum, much smaller than the large scales
$|\vec{P}_{1\perp}|\approx |\vec{P}_{2\perp}|$ characterizing the
overall process, and a special factorization may apply. Let us
first, however, consider the cross section integrated over
$\vec{q}_\perp$. Here, collinear factorization applies, and the
di-jet cross section has the parton-model expression~\cite{Owe87},
\begin{equation}
\frac{d\sigma}{dy_1dy_2dP_\perp^2}=\sum_{ab}x_af_a(x_a)x_bf_b(x_b)
\frac{d\hat\sigma}{d\hat t}(ab\rightarrow cd) \ , \label{lojet}
\end{equation}
where $d\hat\sigma/d\hat t$ is the differential cross section for
the partonic process $ab\rightarrow cd$, with $f_{a,b}$ the
appropriate parton distribution functions. We have defined the
transverse momentum $P_\perp \equiv
|\vec{P}_{1\perp}|=|\vec{P}_{2\perp}|$, and $y_1$ and $y_2$ denote
the rapidities of the two jets. The kinematics are as follows:
\begin{eqnarray}
&&x_a=\frac{P_\perp}{\sqrt{s}}\left(e^{y_1}+e^{y_2}\right),~~~
x_b=\frac{P_\perp}{\sqrt{s}}\left(e^{-y_1}+e^{-y_2}\right)
\nonumber\\
&&\hat s=x_ax_b s,~~~\hat
t=-P_\perp^2\left(e^{y_2-y_1}+1\right),~~~\hat
u=-P_\perp^2\left(e^{y_1-y_2}+1\right) \ .
\end{eqnarray}
Here, $\hat s$, $\hat t$, and $\hat u$ are the usual partonic
Mandelstam variables appearing in the partonic cross sections for
the reactions $ab\rightarrow cd$, namely $\hat{s}=(p_a+p_b)^2$,
$\hat{t}=(p_a-p_c)^2$, $\hat{u}=(p_a-p_d)^2$, in obvious notation
of the partonic momenta. The leading order contributions produce
the di-jet pair exactly balanced, that is back-to-back in the
partonic center-of-mass frame. An imbalance in the transverse
direction is generated by higher-order QCD corrections. At small
but nonzero imbalance between the two jets, the dominant
contributions will come from the intrinsic transverse momenta of
the initial partons. As a model, we will generalize the above
factorization formula to the case of small $\vec{q}_\perp$, in
analogy with the SIDIS and Drell-Yan cases discussed earlier,
taking into account the various contributions to the transverse
momentum dependence coming from the parton distributions and soft
factors:
\begin{eqnarray}
\frac{d\sigma}{dy_1dy_2dP_\perp^2d^2\vec{q}_\perp}&=&
\sum\limits_{ab}\int d^2k_{1\perp}d^2k_{2\perp}d^2\lambda_\perp
x_af_a(x_a,k_{1\perp})x_bf_b(x_b,k_{2\perp})\nonumber\\
&&\times S_{ab\to cd}(\lambda_\perp)H_{ab\to cd}(P_\perp^2)
\delta^{(2)}(\vec{k}_{1\perp}+\vec{k}_{2\perp}+
\vec{\lambda}_\perp-\vec{q}_\perp) \ , \label{jeteq}
\end{eqnarray}
where $S_{ab\to cd}$ is a soft factor for the process $ab\to cd$,
while $H_{ab\to cd}$ is the hard part of the reaction, related to
lowest order to $d\hat\sigma/d\hat t$. We emphasize the overly
simplistic character of Eq.~(\ref{jeteq}) as it stands. The
detailed factorized form (if it exists) will likely be
different; in particular, one expects an interplay of the color
structures of the soft factors and the hard parts, as found in
resummation studies for jet cross sections \cite{Sterman}.

In a similar fashion, we write the Sivers-type contribution
$d\sigma_{TU}$ to the single-polarized cross section,
\begin{eqnarray}
\frac{d\sigma}{dy_1dy_2dP_\perp^2d^2\vec{q}_\perp}&=&d\sigma_{UU}+\vec{e}_z\cdot
\left(\vec{S}_\perp\times \hat{\vec{q}}_\perp \right)d\sigma_{TU}
\ ,
\end{eqnarray}
where $\vec{e}_z$ is the unit vector in the $z$-axis direction. In a
factorized form, we will get:
\begin{eqnarray}
d\sigma_{TU}&=&\sum\limits_{ab}\int
d^2k_{1\perp}d^2k_{2\perp}d^2\lambda_\perp\frac{\vec{k}_{1\perp}\cdot
\hat{\vec{q}}_\perp}{M}
x_aq_{Ta}(x_a,k_{1\perp})x_bf_b(x_b,k_{2\perp})\nonumber\\
&&\times S_{ab\to cd}(\lambda_\perp)H_{ab\to cd}(P_\perp^2)
\delta^{(2)}(\vec{k}_{1\perp}+\vec{k}_{2\perp}+
\vec{\lambda}_\perp-\vec{q}_\perp)  \ .
\end{eqnarray}
We can further simplify the polarized cross section by evaluating
the expression
\begin{eqnarray}
\vec{e}_z\cdot\left(\vec{S}_\perp\times
\hat{\vec{q}}_\perp\right)&=&\frac{|S_\perp|}{|q_\perp|}\vec{e}_z
\cdot\left(\hat{\vec{S}}_\perp\times
(\vec{P}_{1\perp}+\vec{P}_{2\perp})\right)\nonumber\\
&\approx
&\frac{|S_\perp|}{|q_\perp|}|P_\perp|\left(\sin\phi_1+\sin\phi_2\right)
\nonumber\\
&\approx& |S_\perp|\left({\rm
Sgn}(\pi-\theta)\cos\phi_1+\sin\phi_1\frac{|q_\perp|}{2|P_\perp|}\right)
\ ,
\end{eqnarray}
where $\phi_1$ and $\phi_2$ are the azimuthal angles of the two
jets relative to the polarization vector $\vec{S}_\perp$, and
$\theta\equiv\phi_2-\phi_1$ the angle between the two jet transverse
momenta. All these
azimuthal angles are defined in a frame that the polarized proton
is moving the $z$ direction. In the above derivation, we have used
the approximations $|P_{1\perp}|\approx |P_{2\perp}|\approx
|P_\perp|$ and $|q_\perp|\approx |P_\perp||\sin\theta|$, which are
valid at small $q_\perp$ ($\theta$ is close to $\pi$). From the
above result, we can see that there are two terms contributing to
the spin asymmetry: one is with $\cos\phi_1$ and the other with
$\sin\phi_1$. The first term has a Sign function associated, which
gives a positive contribution when $\theta$ is smaller than $\pi$
and a negative one otherwise.

In the above formulas, we did not include any gluon Sivers
function contributions. The gluon Sivers function could dominate
the asymmetry at central rapidities \cite{BoeVog03}. Another
important issue is the relevant Sudakov suppressions for the
asymmetries, which was found to be sizable for the di-jet
correlation in the RHIC energy range \cite{BoeVog03}. In the following
numerical studies, as an order of magnitude estimate for these
asymmetries, we will neglect these effects, which however should
be taken into account in future more detailed studies.

\begin{figure}[t]
\begin{center}
\includegraphics[height=7cm]{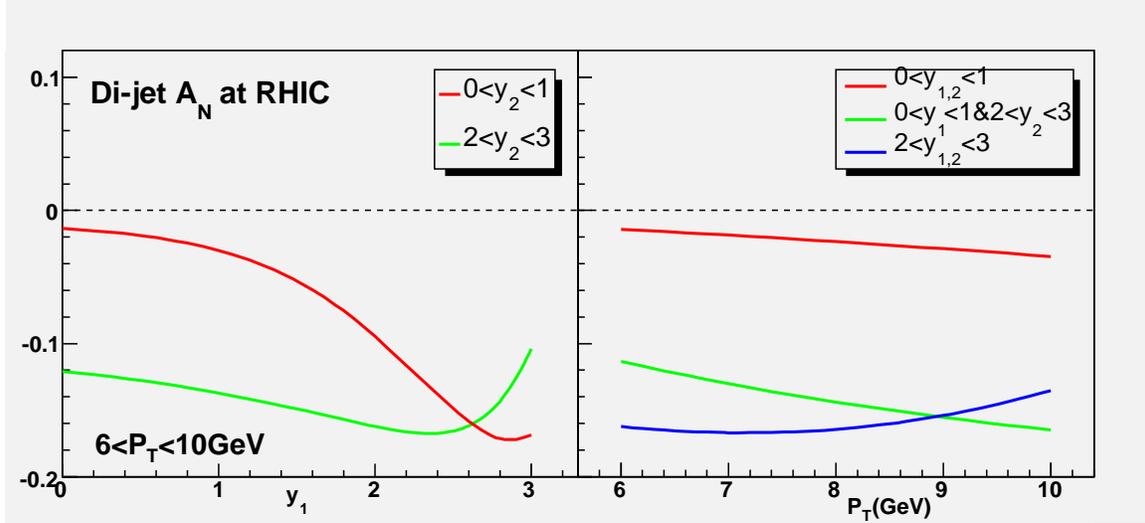}
\end{center}
\vskip -0.4cm \caption{\it  Sivers asymmetries for di-jet
correlations at RHIC, as functions of
rapidity $y_1$ and transverse momentum $P_\perp$. \label{fig12}}
\end{figure}

Following the same procedure that we used for the Sivers
asymmetries in SIDIS and Drell-Yan processes, we can further
simplify the polarized cross section by integrating out the
transverse momentum, but keeping the azimuthal angle dependence
explicit. The differential cross section then can be written as
\begin{eqnarray}
\frac{2\pi
d\sigma}{dy_1dy_2dP_\perp^2d\phi_1}&=&\frac{d\sigma_{UU}}
{dy_1dy_2dP_\perp^2}+\cos\phi_1
\frac{d\sigma_{TU}^{(1)}}{dy_1dy_2dP_\perp^2}+\sin\phi_1
\frac{d\sigma_{TU}^{(2)}}{dy_1dy_2dP_\perp^2}
 \ ,
\end{eqnarray}
where the unpolarized cross section has been given in
Eq.~(\ref{lojet}), and the polarized ones read
\begin{eqnarray}
\frac{d\sigma_{TU}^{(1)}}{dy_1dy_2dP_\perp^2}&=&
\sum_{ab}x_aq_{Ta}^{(1/2)}(x_a)x_bf_b(x_b)
\frac{d\hat\sigma}{d\hat t}(ab\rightarrow cd) \ , \nonumber\\
\frac{d\sigma_{TU}^{(2)}}{dy_1dy_2dP_\perp^2}&=&
\frac{M}{|P_\perp|}\sum_{ab}x_aq_{Ta}^{(1)}(x_a)x_bf_b(x_b)
\frac{d\hat\sigma}{d\hat t}(ab\rightarrow cd) \ , \label{dijetsiv}
\end{eqnarray}
with the distribution $q_T^{(1)}$ defined as
\begin{equation}
q_T^{(1)}(x)=\int d^2k_\perp\frac{k_\perp^2}{2M^2}q_T(x,k_\perp) \
.
\end{equation}
We note that the second term in the polarized cross section is
power-suppressed by $M/P_\perp$. This suppression is due to the
fact that we have integrated over all intrinsic transverse
momentum. Clearly, this term is beyond the approximations we have
made, and we cannot reliably predict it since there will be other
sources of power-suppressed contributions, for example generated
within the Qiu-Sterman mechanism \cite{qiu}. Since it is anyway
expected to be small, we will discard it in the following.
Employing the same set of Sivers functions that we used for our
predictions for the Drell-Yan process above, we then find the
results for $A_N=d\sigma_{TU}^{(1)}/d\sigma_{UU}$ shown in
Fig.~\ref{fig12}. We show the asymmetries as functions of the
rapidity of jet ``1'', and of the jet transverse momentum
$P_\perp$. One can see that the SSA for the di-jet correlation can
become very large, in particular in the forward rapidity region.
Asymmetries of this size should be relatively easily measurable in
the future. We also note that the asymmetry has opposite sign
compared to that for Drell-Yan dimuon production discussed
earlier. The reason for this is that $u$-quark contributions 
dominate in Drell-Yan, thanks to their large electromagnetic charge,
whereas for di-jets, $d$-quark contributions are not
charge-suppressed and in fact dominate, keeping in mind that
the analysis of the HERMES data appears to favor a large $d$ quark
Sivers function. The opposite signs of the Sivers up and down
quark distributions we found in Eq.~(\ref{eq19}) then explains
the opposite signs of the spin asymmetries for Drell-Yan and di-jets.
We note that if the two jets are within the
central rapidity region, our prediction for the asymmetry is much
smaller. As mentioned above, the gluon Sivers function could
dominate the asymmetry in this region \cite{BoeVog03}.

We stress again that if factorization can be shown for the Sivers
SSA in di-jet production, it is likely that the structure of the
resulting expression may differ from the one we use. In
particular, there will be calculable factors that represent the
non-universality of the Sivers functions related to the process
dependence of the gauge links, leading effectively to modified
partonic hard-scattering functions~\cite{BomMulPij05}, at variance
with our use of the standard unpolarized ones. As a test, we have
also used the modified partonic cross sections derived
in~\cite{BomMulPij05}. We find relatively small changes in the
results we obtain. Unfortunately, however, this is not really
representative: the cross sections given in~\cite{BomMulPij05} are
only for the quark-(anti)quark scattering channels, whereas the
dominant contribution in our calculation mostly comes from $qg$
scattering (with the gluon from the unpolarized proton). It
remains to be seen to what extent eventually our predictions will
change, once the process-dependence for the Sivers functions in
di-jet correlations is completely understood.

We finally note that it would also be interesting -- in particular
for measurements with the PHENIX detector -- to study correlations
between hadrons in opposite jets. Such di-hadron correlations
could serve as surrogates for the di-jet correlations we have
discussed above. In this case there will, however, also be
contributions from the Collins mechanism.

\subsection{Jet-Photon Correlations in $p^{\uparrow}p\to {\rm jet}
+\gamma+X$}

\begin{figure}[t]
\begin{center}
\includegraphics[height=7cm]{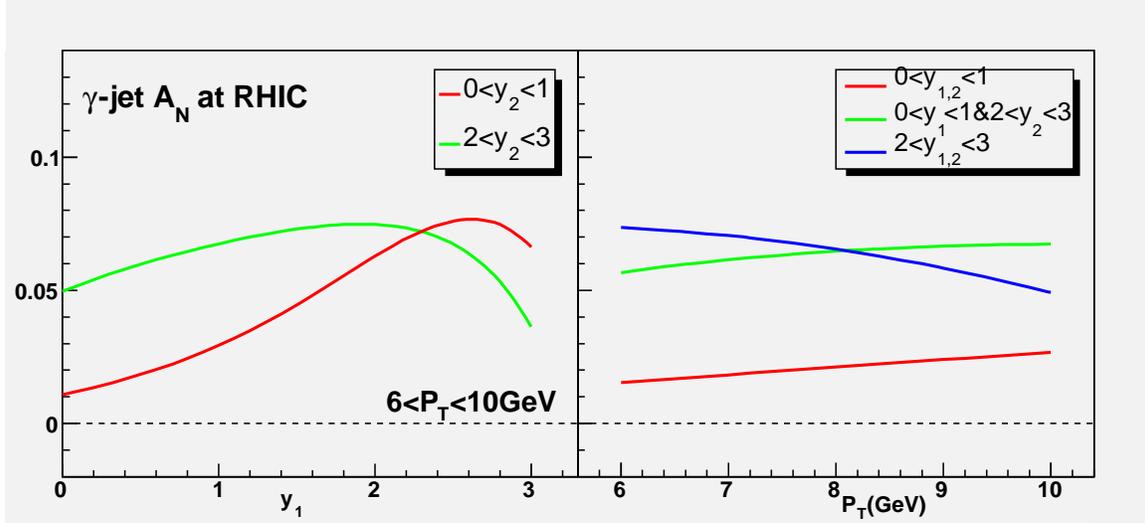}
\end{center}
\vskip -0.4cm \caption{\it Sivers asymmetries for
jet-photon correlations at RHIC, as functions of
photon rapidity $y_1$ and transverse momentum $P_\perp$. \label{fig13}}
\end{figure}

It is straightforward to extend the analysis of di-jet
correlations discussed above to the case of jet-plus-photon
correlations. We simply need to implement the cross sections for
the appropriate Born-level partonic scatterings $q\bar{q}\to\gamma
g$ and $qg\to\gamma q$ in Eq.~(\ref{dijetsiv}). Although events with
a photon suffer from smaller rates than two-jet events, they would
offer additional information on the Sivers functions. It is also
likely that proofs of factorization are more easily obtained here,
since the reactions $q\bar{q}\to\gamma g$ and $qg\to\gamma q$ each
have only a single color structure. Figure~\ref{fig13} shows
results for the single-spin asymmetry for jet-photon correlations
for the same kinematics as for the di-jet case in
Fig.~\ref{fig12}. Variables with the subscript ``1'' denote photon
variables. Again, sizable asymmetries are seen, in particular at
forward rapidities. The asymmetries are somewhat smaller than the
ones we found for di-jets. This is a result of cancellations
between our Sivers $u$ and $d$ functions, due to the larger
weighting factor $4/9$ that the $u$-quark contributions now have
for the prompt-photon case.

\section{Conclusions}

In this paper we have studied single-transverse spin asymmetries
in semi-inclusive deep inelastic scattering and at hadron
colliders. We have analyzed the Sivers and Collins contributions
to the spin asymmetry in SIDIS, and fitted simple
parameterizations of the corresponding functions to recent data
from HERMES. These fits work well and also turn out to be
consistent with COMPASS measurements of the asymmetries in DIS off
a deuteron target. For the Sivers functions, we found dominance of
the down-quark distribution over the up-quark one. The Sivers-$d$
density in SIDIS turns out to be positive, while the $u$-quark
distribution comes out negative. Concerning the Collins functions,
we have found that current data do not yet pin down the relative
size of ``favored'' and ``unfavored'' functions, which is also due
to the fact that the transversity densities are not yet known. We
have also given theoretical arguments that the ``favored'' and
``unfavored'' Collins functions could be of similar size, and of
opposite sign.

We have then investigated Sivers-type single-spin asymmetries at
hadron colliders, focusing on the Drell-Yan process and on di-jet
and jet-photon correlations, all in circumstances where there is a
small measured transverse momentum, but the process is overall
characterized by a large scale. Using the Sivers functions
obtained from the analysis of the HERMES data, we have made
predictions for single-spin asymmetries for these processes. We
find relatively large asymmetries, in particular at forward
rapidities of the observed final state. Such asymmetries should be
measurable with dedicated efforts at RHIC. Besides the additional
valuable information they would give on the Sivers functions and
therefore on the structure of the nucleon, they would also provide
a test of our theoretical understanding of ``naively
time-reversal-odd''  phenomena in QCD. The crucial issues in this
are the factorization of the corresponding cross sections, and the
universality of the Sivers functions, on both of which further
theoretical work is required.

\section*{Acknowledgments}
We thank Elke Aschenauer, Delia Hasch, and Gunar Schnell for providing the
HERMES data, and for discussions regarding the data. We are
grateful to Andrea Bressan for sending us the COMPASS data, and
for discussions. We thank Leslie Bland for stimulating discussions
on jet-photon correlations.  We also thank Harut Avakian, Daniel Boer,
Xiangdong Ji, Naomi Makins, and Matthias Grosse-Perdekamp for
valuable conversations. We also thank Umberto D'Alesio, Rainer
Joosten, and Aram Kotzinian
for correspondence and comments. We are finally grateful to RIKEN,
Brookhaven National Laboratory and the U.S. Department of Energy
(contract number DE-AC02-98CH10886) for providing the facilities
essential for the completion of his work.

{\bf Note Added:} Upon completion of this paper, we noticed the
preprint \cite{umbertonew} where also the Sivers functions were
fitted to the new HERMES and COMPASS data, and predictions for
SSAs in the Drell-Yan process were made. As far as we can see, our
results are in qualitative agreement with those
of~\cite{umbertonew}, keeping in mind that their sign convention
for the Sivers function is opposite to ours and to that in the ``Trento
conventions''.

\end{document}